\documentclass[12pt]{MY_iopart}
\usepackage{amssymb,amsmath}
\usepackage{iopams}
\usepackage{graphics}
\usepackage{epsfig}

\newcommand{\expect}[1]{\langle #1 \rangle}

\newcommand{\GF}{{\rm Green's function}\ }
\newcommand{\GFs}{{\rm Green's functions}\ }
\newcommand{\SE}{{\rm self-energy}\ }
\newcommand{\SEs}{{\rm self-energies}\ }
\newcommand{\NE}{{\rm non-equilibrium}\ }
\newcommand{\MB}{{\rm MB}}
\newcommand{\inter}{{\rm int}}
\newcommand{\summ}{\bullet\hspace{-3.5mm}\sum}
\newcommand{\summline}{\bullet\hspace{-2.9mm}\sum}

\begin{document}

\title[Non-equilibrium many-body transport]
{Many-body current formula and current conservation for non-equilibrium 
fully interacting nanojunctions}

\author{H. Ness, L. K. Dash}
\address{Department of Physics, University of York, Heslington, York YO10 5DD, UK}
\address{European Theoretical Spectroscopy Facility (ETSF)}

\begin{abstract}
  We consider the electron transport properties through fully interacting
  nanoscale junctions beyond the linear-response regime.
  We calculate the 
  current flowing through an interacting region connected to two interacting
  leads, with interaction crossing at the left and right contacts,
  by using a non-equilibrium Green's functions (NEGF) technique.
  Tthe total current at one interface (the left one for
  example) is made of several
  terms which can be regrouped into two sets. The first set corresponds to a 
  very generalised Landauer-like current formula with physical quantities defined only
  in the interacting central region and with renormalised lead self-energies.
  The second set characterises inelastic scattering
  events occuring in the left lead. We show how this term can be negligible 
  or even vanish due to the pseudo-equilibrium statistical properties of the
  lead in the thermodynamic limit.
  The expressions for the different Green's functions
  needed for practical calculations of the current are also provided.
  We determine the constraints imposed by the physical condition
  of current conservation. The corresponding equation imposed on the different
  self-energy quantities arising from the current conservation is derived. We discuss 
  in detail its physical interpretation and its relation with previously
  derived expressions. Finally several important key features are discussed in relation 
  to the implementation of our formalism for calculations of quantum transport in 
  realistic systems.
\end{abstract}
 

\maketitle

\section{Introduction}
\label{sec:intro}

Electronic transport through molecular nanojunctions exhibits many important
new features in comparison with conduction through macroscopic
systems such as bulk or thin layers of semi-conducting molecular crystals as used
in conventional molecular electronics. 
In particular, local interactions, such as Coulomb
interactions between the electrons and scattering from localized
atomic vibrations, become critically important.  
In crude terms, these interactions are more pronounced in nanoscale systems 
because the electronic probability density is concentrated in a small 
region of space; normal screening mechanisms are thus ineffective.  
Developing a theory for the \NE electronic quantum transport through such fully
interacting nanoscale 
junctions is a challenging task, especially when thinking in terms of 
applications for future nanoscale electronics.

Having a simple expression for the electronic current or for the
conductance of a nanoscale object connected to terminals is most
useful.  This is provided by the Landauer formula \cite{Landauer:1970}
in the form of an appealing intuitive physical picture, which
describes the current in terms of local properties of a finite region
(transmission coefficients) and the statistical distribution functions of the
electron reservoirs connected to the central region $C$.  However, in its
original form the Landauer formula deals only with non-interacting
electrons.  This formalism has been used in conjunction with
density-functional theory (DFT) calculations for realistic nanoscale
systems
\cite{Hirose:1994,DiVentra:2000,Taylor:2001,Nardelli:2001,Brandbyge:2002,
  Gutierrez:2002,Frauenheim:2002,Xue:2003,Louis+Palacios:2003,Thygesen:2003,Garcia-suarez:2005}.
It has helped tremendously for the qualitative understanding of the
transport properties of such realistic systems,  
though only on a semi-quantitative level
as the calculated conductance is often one or two orders of magnitude wrong.
The apparent success
of such an approach relies on the fact that DFT maps the many-electron
interacting system onto an effective non-interacting single-particle
Kohn-Sham Hamiltonian suited for the Landauer scattering formalism for transport.
However there are many cases when such a mapping becomes
questionable: for strongly-interacting electron systems, low
dimensional interacting electron system, strongly coupled
electron-phonon systems, to cite only a few.

Furthermore, the single-particle framework of the Landauer approach
cannot be directly transferred to the many-body context by expecting
that a proper inclusion of many-body effects in the single-particle
energy levels will suffice. In fact, it has been shown that
there are many-body corrections to the Landauer formula which cannot
be formulated in terms of single-particle transmission probabilities
\cite{Vignale:2009,Ness:2010}.

When a single-particle-like scheme is still valid even in the presence
of interaction, the Landauer approach can be extended to include
inelastic effects by using inelastic scattering theory in a
generalised Fock-like space
\cite{Orellana:1996,Bonca:1995,Ness:1999,Haule:1999,
  Emberly:2000,Ness:2001,Zitko:2003,Imry:2005}.

In the context of DFT, it has also been shown that the
exchange-correlation part of the interaction that leads to the
presence of an extra $v_{xc}$ potential is actually introducing
corrections to the Landauer-like current. These corrections are
crucial and need to be taken into account when working with DFT
calculations \cite{Koentopp:2006,Koentopp:2008}.

The Landauer formula has been built upon by Meir and Wingreen
\cite{Meir:1992} to extend this formalism to a central scattering
region $C$ containing interactions between particles,
while the left ($L$) and right ($R$) leads are still represented
by non-interacting electron seas.  The current is then expressed
in terms of non-equilibrium Green's functions (NEGF) and self-energies, and in
the most general cases it does not bear any formal resemblance to
the original Landauer formula for the current
\cite{Caroli:1971,Meir:1992,Haug:1996}.  Other generalizations of
Landauer-like approaches to include interactions and inelastic
scattering have been developed, see for example
Refs.~\cite{Imry:2005,Ferretti:2005a,Ferretti:2005b,Vignale:2009}.

However, in real systems the interaction is present throughout the entire system,
even at the nanoscale.  This is even more true for the long-range
Coulomb interaction between bare electrons.  Hence it is
difficult to consider that in the real system all kinds of interaction
will stay localized or will be sufficiently well described by
localized effective interaction in the $C$ region only.
Another reason is that in time-dependent transport, the external field
in the leads may not necessarily be screened
instantaneously. Transient times can be of the same order as the
plasma oscillation (a collective many-body mode of charge oscillation)
period in leads and play an important role in the transient transport
properties of the nanojunction \cite{Myohanen:2010}.
Taking the interaction into the whole system is vital. To achieve this,
the so-called partition-free scheme has been developed
\cite{Cini:1980}.  It allows in principle the calculation of physical
dynamical responses and to include the interactions between the leads
and between the leads and the central region in a quite natural way
\cite{Stefanucci:2004a,Stefanucci:2004b}.

In this paper, we use an alternative approach, and generalize the Meir
and Wingreen formalism (so-called partitioned scheme) to systems where
interaction exists in all the $L,C,R$, as well as at the interfaces
between the three regions.  Since the choice of location of these
interfaces is purely arbitrary and since the interactions exist at and
on both sides of the interface, our approach is equivalent to a
partition-free scheme.

However, while keeping the approach and the NEGF formalism of the
original work of Meir and Wingreen \cite{Meir:1992}, we derive the
most general expression of the current for the fully interacting
system.  From this, we can recover all previously derived transport
expressions or corrections when introducing the appropriate level of
approximation for the interaction.  We also derive and study in detail
the current conservation condition and the constraint that it imposes on
the interaction self-energies.

Our formalism also introduces naturally the generalization of the
concept of the embedding potential when the interaction crosses at the
boundaries. With this new concept and with the condition of current
conservation, we can explore different levels of approximation for
treating the interaction in the different parts of the system, as well
as at the interface. We can then check which approximations are more
suitable for practical numerical calculations of realistic systems.

Although a preliminary account of our formalism is already given in
Ref.~\cite{Ness:2011}, we provide here the full detail of the derivation
and a much more detailed discussion of the physical meaning of our
results as well as of the implementation of the present formalism
for realistic calculations.

The paper is organised as follows: We start with the description of
the system and the notation in Section \ref{sec:model}.  Then we
derive the expression for the current for the fully interacting system
in Section \ref{sec:NEGFtransport}.  
We provide the full derivation of the expressions for the different \GFs 
needed to  calculate the current in Section \ref{sec:differentGFs}.
We derive the conditions imposed by the constraint of current
conservation in Section \ref{sec:currentconserv}.  
We finally conclude and discuss open
questions as well as different schemes to perform the calculation for 
realistic systems in Section \ref{sec:discuss}.
We recall some properties of the \GFs and \SEs in \ref{app:ccGFs}
and B, and provide the proof of an important relation for the current 
conservation is given in \ref{app:gamma_proof}.

\section{System model}
\label{sec:model}

The system consists of two electrodes, labelled $L$ and $R$ for left and right
respectively, which connect a
central region $C$ via a set of coupling matrix elements $V_{LC,RC}$. The
interaction---which we specifically leave undefined
(e.g. electron-electron or electron-phonon)--is assumed to be
well described in terms of the single-particle self-energy $\Sigma^{\rm MB}$ and
spreads over the entire system.

For the calculation of the current, we introduce two interfaces $LC$ and $RC$ 
defining the three regions $L,C$ and $R$ and use different labels 
to name the electronic states on each sides of these interfaces.
The labels $\{\lambda,\lambda'\},\{n,m\},\{\rho,\rho'\}$ are used to represent 
the complete and orthogonal set of states for 
the $L$, $C$ and $R$ regions respectively.

\begin{figure}
  \centering
  \includegraphics[clip,width=0.8\textwidth]{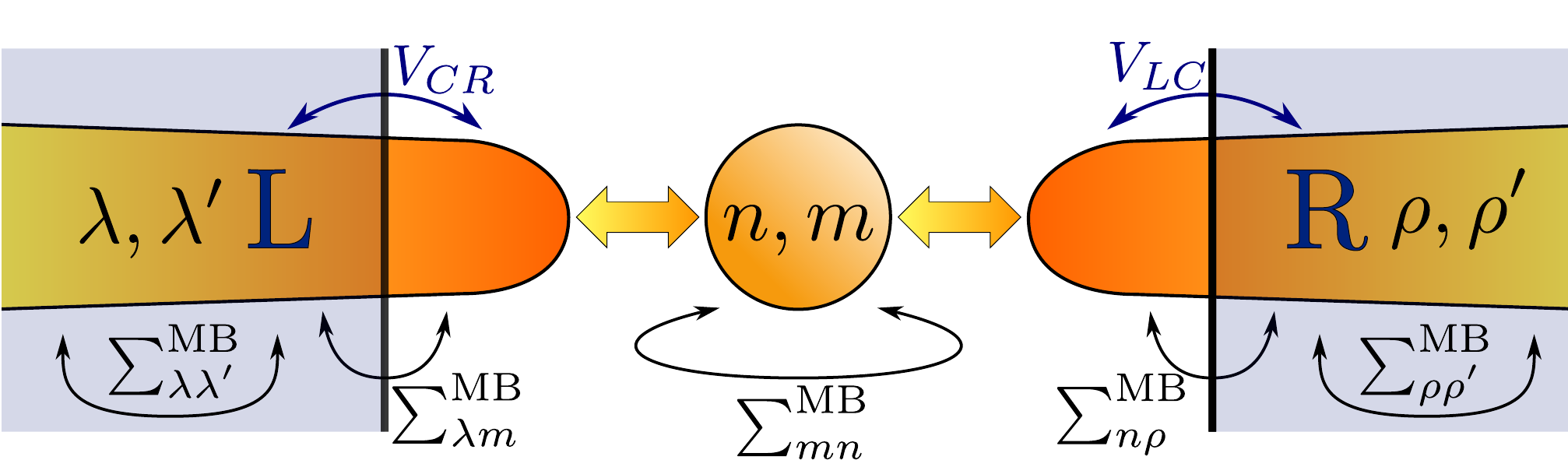}
  \caption{Schematic representation of a central scattering 
region $C$ connected to the left $L$ and right $R$ electrodes, 
with respective quantum-state labels $\{\lambda\},\{n\},\{\rho\}$ for the three $L,C,R$ subspaces.
The electronic coupling of the region $C$ to the $L(R)$ region is given by $V_{LC/CL}$ ($V_{RC/CR}$),
and the many-body interaction is represent by $\Sigma^{\rm MB}$ within all regions as well as
across the $LC$ and $CR$ interfaces. 
  The $LC/RC$ interfaces can be chosen to be located at the contacts
  between the molecule and the leads, or inside the leads (as shown
  above).}
  \label{fig:system}
\end{figure}

In the following, we will use either the full notation or a compact notation for the
Green's functions $G$ and the self-energy $\Sigma$.  Following
Figure~(\ref{fig:system}), the matrix elements of $M_{ij}$ of a \GF or of
a \SE are annotated $M_C$ for the matrix elements of the
central region $M_{nm}$. We use the notation
$M_L$ for $M_{\lambda \lambda'}$ and $M_R$ for $M_{\rho \rho'}$.
The matrix elements are also annotated $M_{LC}$ for $M_{\lambda m}$ or $M_{CL}$ for $M_{n
  \lambda'}$ and $M_{RC}$ for $M_{\rho m}$ or $M_{CR}$ $M_{n \rho'}$.
For  the hopping matrix elements we use $V_{\lambda m}$ or $V_{n \lambda'}$ for $V_{LC}$ or
$V_{CL}$ at the left interface 
and $V_{\rho m}$ or $V_{n \rho'}$ for $V_{RC}$ or $V_{CR}$ at the right interface.

\section{The current formula}
\label{sec:NEGFtransport}

\subsection{Non-equilibrium Green's functions on the Keldysh contour $C_K$}
\label{sec:NEGF-EOM}

The Green's function on the Keldysh contour $C_K$ is defined as 
\begin{equation}
\label{eq:defGFonCK}
G(1,2)	= - {\rm i} \langle \mathcal{T}_{C_K} \Psi(1) \Psi^\dag(2) \rangle,
\end{equation}
where $(1,2)$ stands for a composite index for space
$\mathbf{x}_{1,2}$ (i.e. states $\lambda, n$ or $\rho$ in the $L, C$
or $R$ regions respectively) and time $\tau_{1,2}$ on the time-loop
contour $C_K$. The time ordering $\mathcal{T}_{C_K}$ of the product
of fermion creation ($\Psi^\dag$) and annihilation ($\Psi$) quantum fields
is performed on the time-loop contour $C_K$
\cite{Keldysh:1965,Danielewicz:1984,Chou:1985,vanLeeuwen:2006}.

The \GF obeys the equation of motion on the contour $C_K$ \cite{Danielewicz:1984,vanLeeuwen:2006}:
\begin{equation}
\label{eq:EOM}
[i\partial_{\tau_1} - \bar{h}_0(1)]G(1,1^\prime) =  \delta(1 - 1^\prime) +
\int {\rm d}3 \Sigma(1,3)G(3,1^\prime),
\end{equation}
and the adjoint equation reads
\begin{equation}
\label{eq:EOMadj}
[-i\partial_{\tau_1^\prime} - \bar{h}_0(1^\prime)]G(1,1^\prime) =
\delta(1 - 1^\prime) + \int {\rm d}2 G(1,2) \Sigma(2,1^\prime),
\end{equation}
where $\bar{h}_0(1)$ is the non-interacting Hamiltonian.

\subsection{The current}
\label{sec:current}

Using the continuity equation $\nabla \vec{j}(1) + \partial_t n(1) =
0$, one can write down the current through the interface between the $L$
and $C$ regions. It should be noted that considering an interface
between the $L$ ($R$) and $C$ regions is a merely a virtual partitioning for
mathematical convenience---the interactions are present throughout the entire
system. It is also useful to consider such $LC/RC$ interfaces in
order to connect our results to previously derived expressions within
the partitioning scheme.  The location of these interfaces is also
arbitrary, but for convenience in future numerical computation,
different choices are possible: for example at the contacts between
the leads and the ends of the molecule, or at the contacts between
the leads and the so-called extended
molecule, which already contains part of the leads.

After integration of the continuity equation over the (half) space of
the $L$ region, one obtains the following expression for the current
flowing through the left interface (from $L$ to $C$):
\begin{equation}
\label{eq:defIL}
I_L(t)	= - e \frac{\rm d}{{\rm d}t} \langle \hat{N}_L(t) \rangle,
\end{equation}
where $\langle \hat{N}_L(t) \rangle$ is the total number of electrons
in the L region. It is obtained from the lesser Green's function as
\begin{equation}
\label{eq:defNL}
 \langle \hat{N}_L(t) \rangle = \sum_\lambda -{\rm i} G^<_{\lambda\lambda}(t,t).
\end{equation}

To calculate the time derivative of $G^<_{\lambda\lambda}(t,t)$, we
first go back to the full two-time dependence of $G^<$, and then take
the equal-time limit after performing the derivatives:
\begin{equation}
\label{eq:derGlesser}
  \frac{d}{dt}\expect{\hat{N}_L(t)}   = 
  \sum_\lambda\left( -i
    \frac{d}{dt_1} G^<_{\lambda\lambda}(t_1,t) - i
    \frac{d}{dt_2} G^<_{\lambda\lambda}(t,t_2) \right)_{t_1=t_2= t}
\end{equation}

Using the equations of motion, Eqs.~(\ref{eq:EOM}) and (\ref{eq:EOMadj}),
for the Green's functions on $C_K$, one obtains the current $I_L$ as
\begin{equation}
\label{eq:ILafterEOM}
I_L(t) = \frac{e}{\hbar} {\rm Tr}_\lambda 
\left[ (\Sigma G)^<(t,t) - (G \Sigma)^<(t,t) \right]
\end{equation}
where we have re-introduced the $\hbar$ to have the correct units of current and conductance.

Using the rules of analytical continuation (see \ref{app:analyticalcontinuation}), we get
\begin{equation}
\label{eq:IL_Trace_tt}
  \begin{split}
 {\rm Tr}_\lambda[\ldots] = 
 {\rm Tr}_\lambda \left[ \left(\Sigma^{\MB,<}
     G^a\right) (t,t) + \left( \left( V_{LC} + \Sigma^{\MB,r}\right)(t,t)
     G^< \right) \right. \\ \left. - \left( G^< \left(V_{LC} + \Sigma^{\MB,a} \right)
   \right)(t,t) - \left(G^r\Sigma^{\MB,<}\right)(t,t) \right] .
  \end{split}
\end{equation}
There are no lesser and greater components for $V_{LC}$ since its time
dependence is local, i.e.\ $V_{LC}(t,t') = V_{LC}(t) \delta(t-t')$.

In the steady state, all double-time quantities $X(t,t')$ depend only on the time difference
$X(t-t')$. One obtains the following expression for the current $I_L$ after Fourier transform,
\begin{equation}
\label{eq:ILafterEOM_omega}
I_L = \frac{e}{\hbar} \int \frac{{\rm d}\omega}{2\pi}
{\rm Tr}_\lambda 
\left[ (\Sigma(\omega) G(\omega))^< - (G(\omega) \Sigma(\omega))^< \right]
\end{equation}
or equivalently
\begin{equation}
\label{eq:IL_Trace_w}
  \begin{split}
 I_L = \frac{e}{\hbar} \int \frac{d\omega}{2\pi} {\rm Tr}_\lambda \left[
    V_{LC}G^<(\omega) - G^<(\omega) V_{LC}   + \Sigma^{\MB,<}(\omega) 
   G^a(\omega)   + \Sigma^{\MB,r}(\omega) G^<(\omega) \right. 
\\ \left.-
   G^<(\omega)\Sigma^{\MB,a}(\omega) - G^r(\omega)
   \Sigma^{\MB,<}(\omega) \right].
  \end{split}
\end{equation}

Now that we have sorted out the Keldysh components, we have to
sort out the index (matrix elements) parts.  First we concentrate
on ${\rm Tr}_\lambda \left[ (\Sigma^{\rm MB} G)^< \right]$ and ${\rm
  Tr}_\lambda \left[ (G \Sigma^{\rm MB})^< \right]$.
We thus have 
(we use the symbol $\summline$ for summation to have a better graphical
distinction between the sums and the self-energies $\Sigma$):

\begin{equation}
\label{eq:Trace_LCR_SigmaG}
  \begin{split}
{\rm Tr}_\lambda \left[ (\Sigma^{\MB} G)^< \right]
= 
\summ_{\lambda,n} (\Sigma^{\MB}_{\lambda n} G_{n \lambda})^<
+
\summ_{\lambda,\lambda'} (\Sigma^{\MB}_{\lambda \lambda'} G_{\lambda' \lambda})^<
+
\summ_{\lambda,\rho} (\Sigma^{\MB}_{\lambda \rho} G_{\rho \lambda})^<,
  \end{split}
\end{equation}
and similarly for  ${\rm Tr}_\lambda \left[ (G \Sigma^{\MB})^< \right]$:
\begin{equation}
\label{eq:Trace_LCR_GSigma}
  \begin{split}
{\rm Tr}_\lambda \left[ (G \Sigma^{\MB})^< \right]
=
\summ_{\lambda,n} (G_{\lambda n} \Sigma^{\MB}_{n \lambda})^<
+
\summ_{\lambda,\lambda'} (G_{\lambda \lambda'} \Sigma^{\MB}_{\lambda' \lambda})^<
+
\summ_{\lambda,\rho} (G_{\lambda \rho} \Sigma^{\MB}_{\rho \lambda})^<.
  \end{split}
\end{equation}
In this present version of the theory, we assume that the matrix
elements $\Sigma^{\rm MB}_{\rho \lambda}$ and $\Sigma^{\rm
  MB}_{\lambda \rho}$ vanish, as there is no direct
interaction between the left and right electrode.  
Because of the geometry and the heterogeneity of the nanojunctions and 
because of the different dimensionality of the leads and the central region,
we assume that there is an effective screening of the Coulomb interaction so that 
the electrons of the $L$ and $R$ leads do not
interact directly via any $\Sigma^{\rm MB}_{\rho \lambda}$ or
$\Sigma^{\rm MB}_{\lambda \rho}$ matrix elements. 
The distance $\vert \mathbf{x}_\lambda -\mathbf{x}_\rho \vert$ between two 
points in the $L$ and $R$ leads respectively is large enough so that the 
spatial decay of $\Sigma^{\rm MB}_{\lambda \rho} = \Sigma^{\rm MB}(\vert
\mathbf{x}_\lambda -\mathbf{x}_\rho \vert)$ make the contribution of
$\Sigma^{\rm MB}_{\lambda \rho}$ zero or negligible.
In other words, the presence of the $L$ and $R$ electrodes affect directly the
central region $C$ via $\Sigma^{\MB}_{LC}$ and
$\Sigma^{\MB}_{RC}$. However the $L$ electrode do not affect directly the $R$ 
electrode (and vice-versa), but 
only indirectly via exchange and correlation effects involving the states 
of the central region $C$.
This assumption seems to be valid when the size of the central region is 
of the order of several atoms (i.e.\ a molecule).  
If the central region were to be a single atomic impurity coupled to two 
continuum of delocalised electron state, this assumption would not be valid.

We now turn to the evaluation of the sums in
Eqs.(\ref{eq:Trace_LCR_SigmaG}) and (\ref{eq:Trace_LCR_GSigma}).
First we concentrate on the $\summline_{\lambda,\lambda'}$ sums:
\begin{equation}
  \begin{split}
      \label{eq:sums_lambdas}
      & \summ_{\lambda,\lambda'}  (\Sigma^{\MB}_{\lambda \lambda'} 
      G_{\lambda' \lambda})^< -
      \summ_{\lambda,\lambda'} (G_{\lambda \lambda'} \Sigma^{\MB}_{\lambda' \lambda})^<  \\
      & = \summ_{\lambda,\lambda'} \Sigma^{\MB, <}_{\lambda \lambda'}
      G_{\lambda' \lambda}^a + \Sigma^{\MB,r}_{\lambda \lambda'}
      G_{\lambda' \lambda}^<  
       - G^<_{\lambda \lambda'}
      \Sigma^{\MB,a}_{\lambda' \lambda}
      - G^r_{\lambda \lambda'} \Sigma^{\MB,<}_{\lambda' \lambda} \\
      & = \summ_{\lambda,\lambda'} \Sigma^{\MB, <}_{\lambda \lambda'} (
      G_{\lambda' \lambda}^a - G_{\lambda' \lambda}^r )
      + ( \Sigma^{\MB,r}_{\lambda \lambda'} - \Sigma^{\MB,a}_{\lambda' \lambda}) G_{\lambda' \lambda}^< \\
      & = \summ_{\lambda,\lambda'} \Sigma^{\MB, <}_{\lambda \lambda'}
      (G^<-G^>)_{\lambda' \lambda}
      + ( \Sigma^{\MB,>} - \Sigma^{\MB,<} )_{\lambda \lambda'} G_{\lambda' \lambda}^< \\
      & = \summ_{\lambda,\lambda'} \Sigma^{\MB, >}_{\lambda \lambda'}
      G^<_{\lambda' \lambda}
      - \Sigma^{\MB,<}_{\lambda \lambda'} G_{\lambda' \lambda}^>
      = {\rm Tr}_{\lambda} \left[ \Sigma^{\MB, >}_L G^<_L -
        \Sigma^{\MB,<}_L G^>_L \right].
   \end{split}
\end{equation}
In the second line, we have used the rules of analytical
continuation. In the third, we have used the equivalent of cyclic
permutation in the calculation of a trace, i.e.\ swapping the index
$\lambda$ and $\lambda'$ in the last two terms. This is possible here
since the sums and all matrix elements are defined in the single
subspace of the $L$ electrode.  The final result looks like the
collision terms usually obtained in the derivation of the generalized
Boltzmann equation from quantum kinetic theory.  They correspond to
the particle production (scattering-in) and absorption or hole
production (scattering-out) related to inelastic processes
(i.e.\ non-diagonal elements of the \SE on the time-loop contour
$\Sigma^<$) occurring in the left electrode.
It has also been shown that the integration of such term vanishes as a
result of the gauge degree of freedom \cite{Kita:2010}. In Section \ref{sec:TrII},
we use another route to show how and why these terms can vanish by using
generalised \NE distribution functions.

Now let's concentrate on the $\summline_{\lambda,n}$ sums in
Eq.(\ref{eq:Trace_LCR_SigmaG}) and Eq.(\ref{eq:Trace_LCR_GSigma}).
Once more using the rules of analytical continuation, we can see that
we need the knowledge of the following \GFs matrix elements:
$G^<_{n\lambda}, G^<_{\lambda n}, G^a_{n\lambda}$ and $G^r_{\lambda
  n}$.  For this we use the Dyson-like equation defined for the
non-diagonal elements: $G^<_{n\lambda}=\langle n\vert (G\Sigma
g)^<\vert\lambda\rangle$, We shall not go into the detail of the full
calculations for all four \GF matrix elements; instead we concentrate
on one matrix element $\langle n\vert (G\Sigma
g)^<\vert\lambda\rangle$ to show the mechanism of the derivation:

\begin{equation}
\label{eq:Gless_nlambda}
\begin{split}
G^<_{n\lambda} & =\langle n\vert (G\Sigma g)^<\vert\lambda\rangle
=\langle n\vert G^r \Sigma^< g^r + G^< \Sigma^a g^a + G^r \Sigma^r g^< \vert\lambda\rangle \\
& =\bullet\hspace{-7.5mm}\sum_{\lambda_1,\lambda_2,\rho,m}
  G^r_{n\lambda_1} \Sigma^<_{\lambda_1 \lambda_2} g^a_{\lambda_2 \lambda}
+ G^r_{n\rho} \Sigma^<_{\rho \lambda_2} g^a_{\lambda_2 \lambda} 
+ G^r_{nm} \Sigma^<_{m \lambda_2} g^a_{\lambda_2 \lambda} \\
& \quad\quad + G^<_{n\lambda_1} \Sigma^a_{\lambda_1 \lambda_2} g^a_{\lambda_2 \lambda}
+ G^<_{n\rho} \Sigma^a_{\rho \lambda_2}  g^a_{\lambda_2 \lambda} 
+ G^<_{nm} \Sigma^a_{m \lambda_2} g^a_{\lambda_2 \lambda} \\
& \quad\quad + G^r_{n\lambda_1}\Sigma^r_{\lambda_1 \lambda_2} g^<_{\lambda_2 \lambda}
+ G^r_{n\rho} \Sigma^r_{\rho \lambda_2} g^<_{\lambda_2 \lambda}
+ G^r_{nm}\ \Sigma^r_{m \lambda_2}\ g^<_{\lambda_2 \lambda} ,
\end{split}
\end{equation}
with $\Sigma^{a/r}_{m \lambda}=V_{m \lambda}+\Sigma^{\MB,a/r}_{m
  \lambda}$ and $\Sigma^<_{m \lambda}=\Sigma^{\MB,<}_{m \lambda}$. One has to
keep in mind that we use a model such that $\Sigma^x_{\rho \lambda}=0$.

As we show in detail in Section \ref{sec:renorm_gL}, the interaction defined within the subspace
of the $L$ lead $\Sigma^{a/r}_{\lambda_1 \lambda_2}$ can
be factorized out and included in the renormalized 
\GFs $\tilde g^{a/r,<}_{\lambda_1 \lambda_2}$ of the $L$ lead.
Hence the matrix elements $G^<_{n\lambda}=\langle n\vert (G\Sigma g)^<\vert\lambda\rangle$ can be recast as
$G^<_{n\lambda}=\langle n\vert (G_C\ \Sigma_{CL}\ \tilde{g}_L)^<\vert\lambda\rangle$ with
\begin{equation}
\label{eq:Gless_nlambda_bis}
G^<_{n\lambda}= \summ_{m,\lambda'}
G^r_{nm}\ \Sigma^r_{m \lambda'}\ \tilde{g}^<_{\lambda' \lambda}
+ G^r_{nm}\ \Sigma^{\MB,<}_{m \lambda'}\ \tilde{g}^a_{\lambda'
  \lambda}
+ G^<_{nm}\ \Sigma^a_{m \lambda_2}\ \tilde{g}^a_{\lambda' \lambda} .
\end{equation}

Similarly, we find that 
\begin{equation}
\label{eq:G_nlambda}
\begin{split}
G^<_{\lambda n} & = \langle \lambda\vert
(\tilde{g}_L\ \Sigma_{LC}\ G_C)^<\vert n \rangle , \\
G^a_{n\lambda}  & = \langle n\vert (G_C\ \Sigma_{CL}\
\tilde{g}_L)^a\vert\lambda\rangle , \\ 
G^r_{\lambda n} & = \langle
\lambda\vert (\tilde{g}_L\ \Sigma_{LC}\ G_C)^r\vert n \rangle .
\end{split}
\end{equation}

Finally by using the rules of analytical continuation for the above
matrix elements, we find that the current flowing through the left $L$ interface
Eq.(\ref{eq:IL_Trace_w}) can be recast as
\begin{equation}
     \label{eq:ILfinal}
\begin{split}
 I_L = \frac{e}{\hbar} \int \frac{d\omega}{2\pi} 
& {\rm Tr}_n\left[ G^r_C \tilde{\Upsilon}^{L,l}_C + G^a_C
  (\tilde{\Upsilon}^{L,l}_C)^\dagger + G^<_C \left( \tilde\Upsilon^L_C -
  (\tilde{\Upsilon}^L_C)^\dagger \right) \right] \\
 + & {\rm Tr}_\lambda\left[\Sigma^{\MB,>}_{L} G^<_{L} 
     - \Sigma^{\MB,<}_{L} G^>_{L} \right]
\end{split}
\end{equation}
where
\begin{equation}
\label{eq:Upsilons}
\begin{split}
\tilde\Upsilon^L_C & = \Sigma^a_{CL}\ \tilde{g}^a_L\ \Sigma^r_{LC} , \\
\tilde\Upsilon^{L,l}_C & = \Sigma^<_{CL} \left( \tilde{g}^a_L - \tilde{g}^r_L \right) \Sigma^r_{LC}
+ \Sigma^r_{CL}\ \tilde{g}^<_L\ \Sigma^r_{LC} \\
& = (\Sigma \tilde{g})^<_{CL}\  \Sigma^r_{LC} - \Sigma^<_{CL}\ (\tilde{g} \Sigma)^r_{LC} ,
\end{split}
\end{equation}
and
\begin{equation}
\label{eq:Upsilons_dag}
\begin{split}
(\tilde\Upsilon^L_R)^\dag & = \Sigma^a_{CL}\ \tilde{g}^r_L\ \Sigma^r_{LC} , \\
(\tilde\Upsilon^{L,l}_C)^\dag & = \Sigma^a_{CL} \left( \tilde{g}^a_L - \tilde{g}^r_L \right) \Sigma^<_{LC}
- \Sigma^a_{CL}\ \tilde{g}^<_L\ \Sigma^a_{LC} \\
& = (\Sigma \tilde{g})^a_{CL}\  \Sigma^<_{LC} - \Sigma^a_{CL}\ (\tilde{g} \Sigma)^<_{LC} .
\end{split}
\end{equation}
Eqs.~(\ref{eq:ILfinal}), (\ref{eq:Upsilons}) and (\ref{eq:Upsilons_dag}) are the main results 
of this work for the current formula.
The current $I_L$ flowing at the left $LC$ interface is given by two
traces: the first trace is a generalisation of the Meir and Wingreen
expression of the current to the cases where there are both
interaction within the left electrode and crossing at the $LC$
contact.  The second trace is a term related to inelastic
transport effects involving summation over the left electrode
states/sites only. 

An expression similar to Eq.(\ref{eq:ILfinal}) can be obtained for the current $I_R$ 
flowing at the right $RC$ interface, by swapping the
index $L\leftrightarrow R$ and changing the sign to keep the same convention for positive
current flowing from the left to right direction. We find
\begin{equation}
     \label{eq:IRfinal}
\begin{split}
 I_R = - \frac{e}{\hbar} \int \frac{d\omega}{2\pi} 
& {\rm Tr}_n\left[ G^r_C \tilde{\Upsilon}^{R,l}_C + G^a_C
  (\tilde{\Upsilon}^{R,l}_C)^\dagger + G^<_C \left( \tilde\Upsilon^R_C -
  (\tilde{\Upsilon}^R_C)^\dagger \right) \right] \\
 + & {\rm Tr}_\rho\left[\Sigma^{\MB,>}_{R}
   G^<_{R} -
   \Sigma^{\MB,<}_{R} G^>_{R} \right]
\end{split}
\end{equation}

We now comment on the physical meaning and implication of the new $\Upsilon^{\alpha}_C$
quantities and the different traces entering the definition of the current.

\subsection{Relationships between the $\tilde\Upsilon^L_C$ quantities}
\label{sec:theUpsilons}

From the definition of $\tilde\Upsilon^{L,l}_C$,  
Eq.~(\ref{eq:Upsilons}), we can also define the quantity 
$\tilde\Upsilon^{L,g}_C$ using the greater components $\Sigma^>_{CL}$ and $\tilde{g}^r_L$ 
such as
\begin{equation}
\label{eq:Upsilon_g}
\tilde\Upsilon^{L,g}_{C}  = \Sigma^>_{CL} \left( \tilde{g}^a_L - \tilde{g}^r_L \right) \Sigma^r_{LC}
+ \Sigma^r_{CL}\ \tilde{g}^>_L\ \Sigma^r_{LC} .
\end{equation}
It is now easy to show that $\tilde\Upsilon^l_{LC}$ and
$\tilde\Upsilon^g_{LC}$ are related to each other by
\begin{equation}
\label{eq:Upsilon_relation}
\tilde\Upsilon^{L,g}_{C}  - \tilde\Upsilon^{L,l}_{C} = 
(\tilde\Upsilon^L_C)^\dag - \tilde\Upsilon^L_C . 
\end{equation}
This is a very interesting relationship in the sense that the $\tilde\Upsilon^L_C$
quantities obey a relation of the type: (greater) $-$ (lesser)=(retarded) $-$ (advanced)
as for conventional \SEs or Green's functions, though $\tilde\Upsilon^{L,l}_C$ and $\tilde\Upsilon^{L,g}_C$
are not proper lesser and greater quantities.
It should also be noted that the
different $\tilde\Upsilon^L_C(\omega)$ play a similar role as the
the lead self-energies (or embedding
potentials \cite{Inglesfield:1981,Inglesfield:2005,Inglesfield:2008}), defined as
$\Sigma^{L,x}_C(\omega) = V_{CL}\ g^x_L(\omega)\ V_{LC}$ when the interactions
are not crossing at the contacts.
However, they are not simply related to the straightforward generalization of
these embedding potentials.
The latter have the following form (see also the expression for the \GFs given 
in Section \ref{sec:differentGFs}) 
\begin{equation}
\label{eq:Yx_L}
\tilde Y^{L,x}_C(\omega) = (\Sigma_{CL}(\omega)\ \tilde g_L(\omega)\ \Sigma_{LC}(\omega))^x 
\end{equation}
with $\Sigma_{LC/CL}(\omega)=V_{LC/CL}+\Sigma^\MB_{LC/CL}(\omega)$
and $x=(>,<,r,a)$. 
The
rules of analytical continuation for $\tilde Y^{L,<}_C$ and $\tilde Y^{L,r}_C$ do not
give the same expression as for $\tilde\Upsilon^{L,l}_C$ or
$\tilde\Upsilon^L_C$.  For example $\tilde Y^{L,r}_C= \Sigma^r_{CL}\ \tilde g^r_L\
\Sigma^r_{LC} \ne \tilde\Upsilon^L_C$ and 
$\tilde Y^{L,<}_C \ne \tilde\Upsilon^{L,l}_C$.

This leads us to an important proof of this work:
Because of ({\it a}) the very existence of the interaction crossing at the contact, 
of ({\it b}) the fact that $\Sigma^a_{L\alpha /
  \alpha C} \ne \Sigma^r_{L\alpha / \alpha C}$ (as opposed to the
non-interacting case where $V^a_{L\alpha/\alpha C} = V^r_{L\alpha /
  \alpha C} = V_{L\alpha / \alpha C}$, 
and of ({\it c}) the rules of analytical continuation for products of three quantities, 
the usual cyclic
permutation used in the calculation of the trace ${\rm
  Tr}_\lambda[(\Sigma G)^<-(G \Sigma)^<]$ cannot be used to transform
the initial trace over $\{\lambda\}$ onto a trace over $\{n\}$ only.
Therefore the current $I_L$ at the $LC$ contact cannot be expressed simply in
terms of the generalized embedding potentials $\tilde Y^{L,x}_C$, hence the introduction
of the $\tilde\Upsilon^L_C$ quantities.
The current is \emph{not} obtained from a straightforward
generalisation of the Meir and Wingreen formula using the embedding potentials $\tilde Y^{L,x}_C$:
\begin{equation}
  I_L \ne \frac{e}{\hbar} \int \frac{d\omega}{2\pi} {\rm Tr}_n [Y^{L,<}_C G^>_C - Y^{L,>}_C G^<_C] 
  + {\rm Tr}_\lambda [\Sigma^{\MB >}_L G^<_L - \Sigma^{\MB <}_L
  G^>_L].
\end{equation}

\subsection{Non-equilibrium distribution functions in the leads}
\label{sec:TrII}

Now we consider the terms ${\rm Tr}_{\lambda} [ \Sigma^{\MB, >}_L
G^<_L - \Sigma^{\MB,<}_L G^>_L ]$, and show in which conditions this
trace vanishes.  For this we introduce the \NE distribution functions
$f^<_L(\omega)$ and $f^{\MB,<}_L(\omega)$ defined from the generalised
Kadanoff-Baym ansatz \cite{Lipavski:1986} as follows:
\begin{equation}
\label{eq:def_NEdistrib}
\begin{split}
  G^<_L(\omega) & = f^<_L(\omega) G^a_L(\omega) - G^r_L(\omega) f^<_L(\omega) \\
  \Sigma^{\MB,<}_L(\omega) & = f^{\inter,<}_L(\omega)
  \Sigma^{\MB,a}_L(\omega) - \Sigma^{\MB,r}_L(\omega)
  f^{\inter,<}_L(\omega) ,
\end{split}
\end{equation}
and similarly for $f^>_L$ and $f^{\MB,>}_L$, obtained from $G^>_L$ and $\Sigma^{\MB,>}_L$.  
These distribution
functions follow the conditions $f^>_L=f^<_L - 1_L$ and
$f^{\inter,<}_L=f^{\inter,<}_L - 1_L$ (where the identity matrix in the $L$
region is $1_L=\delta_{\lambda \lambda'}$ so that the usual
relationships between the different \GFs (and \SEs) $X^r-X^a=X^>-X^<$
still hold.

For convenience, we consider for the moment that the \NE
distribution functions are diagonal in the corresponding subspace,
i.e. $f^<_{\lambda \lambda'} = f^<_\lambda \delta_{\lambda
  \lambda'}$. However there is no formal difficulty to deal with a
full, non-diagonal, dependence of these density-matrix-like distribution
functions.

Introducing the definition of the \NE distribution functions in
Eq.(\ref{eq:sums_lambdas}), we end up, after
lengthy (but rather trivial) calculations, with
\begin{equation}
  \label{eq:trace_lambda} {\rm Tr}_{\lambda} \left[ \Sigma^{\MB, >}_L
    G^<_L - \Sigma^{\MB,<}_L G^>_L \right]  = (2\pi)^2 {\rm
    Tr}_\lambda \left[(f^<_L - f^{\inter,<}_L) A^\Sigma_L(\omega)
    A^G_L(\omega) \right]
\end{equation}
where the respective spectral functions are obtained from
\begin{equation}
\label{eq:def_spectralfncs2}
 2\pi{\rm i} A^X_L(\omega)   = X^a_L(\omega) - X^r_L(\omega),
\end{equation}
with $X(\omega)=G(\omega)$ or $\Sigma(\omega)$.

Eq.(\ref{eq:trace_lambda}) shows that the collision term is a measure
of the deviation between the two distribution functions
$f^<_L(\omega)$ and $f^{\inter,<}_L(\omega)$.  At equilibrium, because
all distribution functions are equal to the Fermi distribution
$f^<_L=f^{\inter,<}_L=f^{\rm eq}$, this collision term vanishes. At
\NE this is not generally the case.

One can now imagine the following case: the indices $\lambda$
represent a spatial location or a localised electronic state on a
lattice point in the left electrode. For $\lambda$ located well inside
the electrode, the system is in its local equilibrium (the Fermi
distribution with a Fermi level shifted by the left bias) and both
distribution functions $f^<_L$ and $f^{\inter,<}_L$ are equal to 
the left Fermi distribution,
hence these indexes do not contribute to the trace. Only the lattice
points (or states) that are not in local quasi-equilibrium, i.e.\ those
close enough to the central region to experience the potential drop
and the interaction effects at the contact and beyond will contribute
to the trace.
From a computational point of view, this is good news in the sense
that one is not be obliged to perform the summation in ${\rm
  Tr}_{\lambda} [...]$ over all the infinite $\lambda$ indexes of the
semi-infinite left electrode. Only the states/sites for which
$f^<_\lambda - f^{\inter,<}_\lambda \ne 0$ will contribute to the trace.
So if we choose the location of the $LC$ interface far enough inside the
left electrode of the real system, then ${\rm Tr}_{\lambda}
[...]=0$. 
In this sense, we have just provided a formal proof of the
concept of the so-called extended molecule \cite{Strange:2011,
Louis+Palacios:2003,Xue:2003,Brandbyge:2002,Palacios:2002,Palacios:2001}
The extended molecule represents the central region $C$ and consists
of the molecule itself but also a part of the left and right electrodes
to which the molecule is connected. The concept of the extended molecule
has been introduced empirically in realistic calculations of
molecular junctions in order to avoid any problems with the asymptotic
behaviour of the electrostatic potential in the bias of finite (not small)
applied bias.

The value of ${\rm Tr}_{\lambda} [...]$ can also be understood 
as a measure of the ``efficiency'' of the location of the $LC$ interface in the
$L$ electrode. The larger (smaller) the value is, the farther (closer) from local 
equilibrium the $LC$ interface is. This measure can help in finding a good compromise 
between having a sufficiently large extended molecule that is nonetheless small enough 
for tractable numerical calculations.

\subsection{The Meir and Wingreen current formula}
\label{sec:diffapprox}

Using different approximations (for example single-particle approaches, mean-field theories, 
interactions localised in the central region only) for the different interacting self-energies,
we can recover from Eq.~(\ref{eq:ILfinal}) all the previously derived current expressions
for \NE nanojunctions. We have analysed all these connections in detail in Ref.~\cite{Ness:2011} 
and we will not repeat the analysis here.

However, as it will be useful below, we now briefly recall that
Eq.~(\ref{eq:ILfinal}) bears some resemblance to the current expression derived by Meir
and Wingreen \cite{Meir:1992} in the case of interaction present in the central region only:
\begin{equation}
\label{eq:IL_MeirWingreen}
I_L^{\rm MW} = \frac{{\rm i} e}{\hbar} \int \frac{{\rm d}\omega}{2\pi}\
{\rm Tr}_{n} \left[ f_L (G^r_C - G^a_C) \Gamma_C^L + G^<_C \Gamma_C^L \right] .
\end{equation}
The connection becomes more apparent 
when we use the definitions
${\rm i} f_L \Gamma_C^L=V_{CL}\ {g}^<_L\ V_{LC} = \Sigma^{L,<}_C = - (\Sigma^{L,<}_C)^\dag$
and
${\rm i}\Gamma_C^L=V_{CL} ({g}^a_L-g^r_L) V_{LC} = \Sigma^{L,a}_C - \Sigma^{L,r}_C$. 
Hence $I_L^{\rm MW}$ becomes
\begin{equation}
\label{eq:IL_MeirWingreen_bis}
\begin{split}
I_L^{\rm MW} & = \frac{e}{\hbar} \int \frac{{\rm d}\omega}{2\pi} 
 {\rm Tr}_{n} \left[ G^r_C \Sigma^{L,<}_C + G^a_C (\Sigma^{L,<}_C)^\dag  
+ G^<_C  (\Sigma^{L,a}_C - \Sigma^{L,r}_C) \right] .
\end{split}
\end{equation}

One can see by comparing Eq.~(\ref{eq:ILfinal}) and Eq.~(\ref{eq:IL_MeirWingreen_bis})
that the quantities $\tilde\Upsilon^L_C$, $(\tilde\Upsilon^L_C)^\dag$
and $\tilde\Upsilon^{L,l}_C$ 
play the role of the $L$ lead self-energies  
$\Sigma^{L,a}_C$, $\Sigma^{L,r}_C$ and $\Sigma^{L,<}_C$ respectively
in the cases where the interactions cross at the $L$ interface.
However, as we have discussed above, the quantities $\tilde\Upsilon^L_C$ are the forward
generalisation of the embedding potentials $\tilde Y^{L,x}_C$.
Note that, in the model of Meir and Wingreen, the leads are non-interacting, hence the second 
trace  ${\rm Tr}_\lambda\left[...\right]$ in Eq.~(\ref{eq:ILfinal}) vanishes.

\section{The different Green's functions needed}
\label{sec:differentGFs}

For the evaluation of the currents $I_L$ and $I_R$, we need to know
the following \GFs: $G^{a/r,<}_C$ and $G^{a/r,<}_{L,R}$. For this, we
calculate the matrix elements: $G^x_C=\langle n\vert (g+g\Sigma
G)^x\vert m\rangle$, $G^x_L=\langle \lambda\vert (g+g\Sigma G)^x\vert
\lambda' \rangle$, $G^x_R=\langle \rho \vert (g+g\Sigma G)^x\vert
\rho' \rangle$. The results are given in the following subsections.

\subsection{Renormalisation of the Green's functions in the electrodes}
\label{sec:renorm_gL}

First we show that the terms in $\Sigma^{a/r}_{L}$ can
be factorized out and included within the renormalization of the 
left lead \GFs $g^{a/r,<}_{L}$. Hence the matrix elements 
$G^<_{n\lambda}=\langle n\vert (G\Sigma g)^<\vert\lambda\rangle$ 
can be recast as
$G^<_{n\lambda}=\langle n\vert (G_C\ \Sigma_{CL}\ \tilde{g}_L)^<\vert\lambda\rangle$ 
with $\tilde{g}_L$ the renormalised $L$ lead \GF whose definition is given below.

We start from
\begin{equation}
 \label{eq:app_Gless_nlambda_1}
\begin{split}
      G^<_{n\lambda} & =\langle n\vert (G\Sigma g)^<\vert\lambda\rangle \\
      & =\bullet\hspace{-6mm}\sum_{\lambda_1,\lambda_2,m} 
      G^r_{n\lambda_1}\ \Sigma^<_{\lambda_1 \lambda_2}\ g^a_{\lambda_2
        \lambda}
      + G^r_{nm}\ \Sigma^<_{m \lambda_2}\ g^a_{\lambda_2 \lambda} 
      +\  G^<_{n\lambda_1}\ \Sigma^a_{\lambda_1 \lambda_2}\
      g^a_{\lambda_2 \lambda}
      + G^<_{nm}\ \Sigma^a_{m \lambda_2}\ g^a_{\lambda_2 \lambda} \\
      & + G^r_{n\lambda_1}\ \Sigma^r_{\lambda_1 \lambda_2}\
      g^<_{\lambda_2 \lambda} + G^r_{nm}\ \Sigma^r_{m \lambda_2}\
      g^<_{\lambda_2 \lambda} ,
\end{split}
\end{equation}
  hence (we now use Einstein notation for the sums)
\begin{equation}
    \label{eq:app_Gless_nlambda_2}
\begin{split}
          G^<_{n\lambda} - G^<_{n\lambda_1}\ \Sigma^a_{\lambda_1
        \lambda_2}\ g^a_{\lambda_2 \lambda}  = G^r_{n\lambda_1}\
      \Sigma^<_{\lambda_1 \lambda_2}\ g^a_{\lambda_2 \lambda} +
      G^r_{nm}\ \Sigma^<_{m \lambda_2}\ g^a_{\lambda_2 \lambda} +
      G^<_{nm}\ \Sigma^a_{m \lambda_2}\ g^a_{\lambda_2 \lambda}  \\
      + G^r_{n\lambda_1}\ \Sigma^r_{\lambda_1 \lambda_2}\
      g^<_{\lambda_2 \lambda} + G^r_{nm}\ \Sigma^r_{m \lambda_2}\
      g^<_{\lambda_2 \lambda} ,
\end{split}
\end{equation}
  and
\begin{equation}
    \label{eq:app_Gless_nlambda_3}
\begin{split}
      G^<_{n\lambda_1} (1 - \Sigma^a\ g^a)_{\lambda_1 \lambda}  =
      G^r_{n\lambda_1}\ \Sigma^<_{\lambda_1 \lambda_2}\ g^a_{\lambda_2
        \lambda}
      + G^r_{nm}\ \Sigma^<_{m \lambda_2}\ g^a_{\lambda_2 \lambda} \\
       + G^<_{nm}\ \Sigma^a_{m \lambda_2}\ g^a_{\lambda_2 \lambda} +
      G^r_{n\lambda_1}\ \Sigma^r_{\lambda_1 \lambda_2}\ g^<_{\lambda_2
        \lambda} + G^r_{nm}\ \Sigma^r_{m \lambda_2}\ g^<_{\lambda_2
        \lambda} ,
\end{split}
\end{equation}
  so
\begin{equation}
    \label{eq:app_Gless_nlambda_4}
\begin{split}
          G^<_{n\lambda}  = G^r_{n\lambda_1}\ \Sigma^<_{\lambda_1
        \lambda_2}\ \tilde{g}^a_{\lambda_2 \lambda} + G^r_{nm}\
      \Sigma^<_{m \lambda_2}\ \tilde{g}^a_{\lambda_2 \lambda}
      + G^<_{nm}\ \Sigma^a_{m \lambda_2}\ \tilde{g}^a_{\lambda_2 \lambda} \\
       + G^r_{n\lambda_1}\ \Sigma^r_{\lambda_1 \lambda_2}\
      g^<_{\lambda_2 \lambda_1} (1 - \Sigma^a\ g^a)^{-1}_{\lambda_1
        \lambda} + G^r_{nm}\ \Sigma^r_{m \lambda_2}\ g^<_{\lambda_2
        \lambda_1} (1 - \Sigma^a\ g^a)^{-1}_{\lambda_1 \lambda} ,
\end{split}
\end{equation}
  where we define the renormalised \GFs $\tilde{g}^a_L$ for the left $L$
  electrode as
  \begin{equation}
    \label{eq:def_tildega}
    g^a_{\lambda \lambda'} \left( 1- \Sigma^{\MB,a} g^a \right)^{-1}_{\lambda' \lambda_1}
    = \tilde{g}^a_{\lambda \lambda_1} .
  \end{equation}
To solve for Eq.(\ref{eq:app_Gless_nlambda_4}) we also need to know
the matrix element $G^r_{n\lambda_1}$
\begin{equation}
\label{eq:app_Gr_nlambda_1}
G^r_{n\lambda_1}=\langle n\vert (G\Sigma g)^r\vert\lambda_1\rangle
=\langle n\vert G^r \Sigma^r g^r \vert\lambda_1\rangle  =
G^r_{n\lambda_2}\ \Sigma^r_{\lambda_2 \lambda_3}\ {g}^r_{\lambda_3 \lambda_1}
+ G^r_{nm}\ \Sigma^r_{m \lambda_3}\ {g}^r_{\lambda_3 \lambda_1} ,
\end{equation}
hence
\begin{equation}
\label{eq:app_Gr_nlambda_2}
G^r_{n\lambda_3} (1 -\Sigma^r_L\ {g}^r_L)_{\lambda_3 \lambda_1} = 
G^r_{nm}\ \Sigma^r_{m \lambda_3}\ {g}^r_{\lambda_3 \lambda_1} ,
\end{equation}
so
\begin{equation}
\label{eq:app_Gr_nlambda_3}
G^r_{n\lambda_1}  = G^r_{nm}\ \Sigma^r_{m \lambda_2}\ \tilde{g}^r_{\lambda_2 \lambda_1} ,
\end{equation}
with a similar definition for the renormalised \GF $\tilde{g}^r_L$ as
given for $\tilde{g}^a_L$ in Eq.(\ref{eq:def_tildega}).
With our compact notation, we have
\begin{equation}
\label{eq:def_tildeg}
\tilde{g}^{r/a}_L =  {g}^{r/a}_L + {g}^{r/a}_L\ \Sigma^{\MB,r/a}_L\ \tilde{g}^{r/a}_L .
\end{equation}
Using the result of Eq.(\ref{eq:app_Gr_nlambda_3}) into
Eq.(\ref{eq:app_Gless_nlambda_4}) and using the fact that $(1 -
\Sigma^a_L\ g^a_L)^{-1} = (1 + \Sigma^a_L\ \tilde{g}^a_L)$, we find
that
\begin{equation}
\label{eq:app_Gless_nlambda_5}
\begin{split}
G^<_{n\lambda}  & = 
G^r_{nm}\ \Sigma^r_{m \lambda_2}\ \tilde{g}^r_{\lambda_2 \lambda_1}\ \Sigma^<_{\lambda_1 \lambda_2}\ 
\tilde{g}^a_{\lambda_2 \lambda}
+ G^r_{nm}\ \Sigma^<_{m \lambda_2}\ \tilde{g}^a_{\lambda_2 \lambda} 
+ G^<_{nm}\ \Sigma^a_{m \lambda_2}\ \tilde{g}^a_{\lambda_2 \lambda} \\
& +  G^r_{nm}\ \Sigma^r_{m \lambda_2}\ \tilde{g}^r_{\lambda_2 \lambda_1}\ \Sigma^r_{\lambda_1 \lambda_2}\ g^<_{\lambda_2 \lambda_1} (1 +  \Sigma^a\ \tilde{g}^a)_{\lambda_1 \lambda}
+ G^r_{nm}\ \Sigma^r_{m \lambda_2}\ g^<_{\lambda_2 \lambda_1} (1 +  \Sigma^a\ g^a)_{\lambda_1 \lambda} \\ 
& =
G^r_{nm}\ \Sigma^r_{m \lambda_2}\ \tilde{g}^<_{\lambda_2 \lambda} 
+ G^r_{nm}\ \Sigma^<_{m \lambda_2}\ \tilde{g}^a_{\lambda_2 \lambda} 
+ G^<_{nm}\ \Sigma^a_{m \lambda_2}\ \tilde{g}^a_{\lambda_2 \lambda} \\
& = \langle n\vert (G_C\ \Sigma_{CL}\ \tilde{g}_L)^<\vert\lambda\rangle ,
\end{split}
\end{equation}
with
\begin{equation}
\label{eq:def_tildeglesser}
\tilde{g}^<_L = \tilde{g}^{r}_L\ \Sigma^{\MB,<}_L\ \tilde{g}^{a}_L
+ ( 1 + \tilde{g}^{r}_L\ \Sigma^{\MB,r}_L ) {g}^<_L ( 1 + \Sigma^{\MB,a}_L\ \tilde{g}^{a}_L).
\end{equation}

In other words: the $\tilde{g}^x_L$ are the \GF in the left $L$ electrode 
renormalised by the many-body self-energy $\Sigma^{\MB,x}_L$ defined in the same 
subspace of the left $L$ electrode.
Similar results can be derived for the Green's functions of the right $R$ electrode.

\subsection{The retarded Green's function $G^r_C$ in the central region}

We find for $G^r_C = \langle n\vert G^r \vert m\rangle$

\begin{equation}
\label{eq:GrC}
\begin{split}
G^r_C & = {g}^r_C + {g}^r_C\ \Sigma^{\MB,r}_C\ G^r_C + {g}^r_C\ \tilde Y^{L+R,r}_C\ G^r_C \\
      & = \left[ ({g}^r_C)^{-1} - \Sigma^{\MB,r}_C -  \tilde Y^{L+R,r}_C
      \right]^{-1} \\
      &  = \left[ (\tilde{g}^r_C)^{-1} -  \tilde Y^{L+R,r}_C \right]^{-1} ,
\end{split}
\end{equation}
where $\tilde Y^{L+R,r}_C$ is the sum $\tilde Y^{L+R,r}_C=\tilde Y^{L,r}_C+\tilde Y^{R,r}_C$
of the generalised lead
self-energies $\tilde Y^{\alpha,r}_C$ ($\alpha=L,R$) defined previously as
$\tilde Y^{\alpha,r}_C = (\Sigma_{C\alpha} \tilde g_{\alpha} \Sigma_{\alpha C})^r = 
\Sigma^r_{C\alpha} \tilde g^r_{\alpha}\Sigma^r_{\alpha C}$ with
$\Sigma_{C\alpha} = V_{C\alpha} + \Sigma^\MB_{C\alpha}$.

Similar expression can be derived for the advanced Green's function $G^a_C$ 
in the central region by swapping $r \leftrightarrow a$.

\subsection{The lesser Green's function $G^<_C$ in the central region}

We find for $G^r_C = \langle n\vert G^< \vert m\rangle$ that
\begin{equation}
\label{eq:GlessC}
G^<_C = G^r_C\ \left( \Sigma^{\MB,<}_C + \tilde Y^{L+R,<}_C \right) G^a_C,
\end{equation}
with 
$\tilde Y^{L+R,<}_C(\omega) = \tilde Y^{L,<}_C(\omega) + \tilde Y^{R,<}_C(\omega) =
\bullet\hspace{-3mm}\sum_{\alpha=L,R} 
\left( \Sigma_{C\alpha}(\omega) \tilde g_{\alpha}(\omega) \Sigma_{\alpha C}(\omega) \right)^<$.
Using the rules of analytical continuation for products given in \ref{app:analyticalcontinuation},
one gets 
$\tilde Y^{L+R,<}_C=\bullet\hspace{-3mm}\sum_{\alpha=L,R} 
\Sigma^r_{C\alpha} \tilde g^r_{\alpha} \Sigma^<_{\alpha C} 
+ \Sigma^r_{C\alpha} \tilde g^<_{\alpha} \Sigma^a_{\alpha C}
+ \Sigma^<_{C\alpha} \tilde g^a_{\alpha} \Sigma^a_{\alpha C}$.

\subsection{The retarded Green's function $G^r_{L,R}$ in the $L$ and $R$ lead}

We find for $G^r_R = \langle \rho \vert G^r \vert \rho' \rangle$ that

\begin{equation}
\label{eq:GrR}
G^r_{\rho \rho'} = \tilde{g}^r_{\rho \rho'} 
+ \tilde{g}^r_{\rho \rho_1}\ \tilde Y^{CL,r}_{\rho_1\rho_2} G^r_{\rho_2 \rho'} 
\end{equation}
where $\tilde{ Y}^{CL,r}_R$ is an embedding potential of the
effects of the central region $C$ (connected to the left lead) on the right lead. 
It is defined as
\begin{equation}
\label{eq:Cregion_embed_pot1}
\tilde Y^{CL,r}_{\rho_1\rho_2}(\omega) 
= \Sigma^r_{\rho_1 m}(\omega)\ 
\left[ [\tilde{g}^r_C(\omega)]^{-1}  - \tilde Y^{L,r}_C(\omega) \right]^{-1}_{ml}
 \Sigma^r_{l \rho_2}(\omega) ,
\end{equation}
with $\Sigma^r_{RC}=V_{RC}+\Sigma^{\MB,r}_{RC}$ and similarly for
$\Sigma^r_{CR}$.  And 
$[ [\tilde{g}^r_C(\omega)]^{-1}  - \tilde Y^{L,r}_C(\omega) ]^{-1}$ is a retarded \GF 
of the central region renormalized by the interaction inside $C$ and
by the lead self-energy / embedding potential $\tilde Y^{L,r}_C$ of the left lead only,
with $\tilde{g}^r_C$ defined in Eq.(\ref{eq:GrC}) as 
$(\tilde{g}^r_C)^{-1} = ({g}^r_C)^{-1} - \Sigma^{\MB,r}_C$.

Similarly we can find the expressions for the $L$ lead \GF:
\begin{equation}
\label{eq:GrL}
\begin{split}
G^r_L & = \tilde{g}^r_L + \tilde{g}^r_L \tilde Y^{CR,r}_L G^r_L , \\
\tilde Y^{CR,r}_L & = 
\Sigma^r_{LC} \left[ (\tilde{g}^r_C)^{-1}  - \tilde Y^{R,r}_C \right]^{-1}  \Sigma^r_{CL} .
\end{split}
\end{equation}

Finally all the expressions given in this section hold for the advanced $G^a_{L,R}$ by 
swapping $r \leftrightarrow a$.

\subsection{The lesser Green's function $G^<_{L,R}$ in the leads}

We first concentrate on $G^<_L = (G \Sigma g + g)^<_{LL}$.
The calculations are rather lengthy, but somehow trival, and include the 
derivation of many intermediate \GFs and many re-factorizations. We find, 
in agreement with the results of the previous section, that
 
\begin{equation}
\label{eq:GlesserL}
G^<_L  = (1 + G^r_L \tilde Y^{CR,r}_L ) \tilde{g}^<_L (1 + \tilde Y^{CR,a}_L G^a_L)
+ G^r_L \tilde Y^{CR,<}_L G^a_L  ,
\end{equation}
and
\begin{equation}
\label{eq:GlesserL_plus}
\begin{split}
\tilde Y^{CR,x}_L   & = (\Sigma_{LC} \tilde{g}^R_C  \Sigma_{CL} )^x , \\
\tilde{g}^{R,r/a}_C & = \left[ (\tilde{g}^{r/a}_C)^{-1}  - \tilde Y^{R,r/a}_C \right]^{-1} , \\
\tilde{g}^{R,<}_C   & = (1 + \tilde{g}^{R,r/a}_C \tilde Y^{R,r}_C ) \tilde{g}^<_C 
			(1 + \tilde Y^{R,a}_C \tilde{g}^{R,a}_C)
+ \tilde{g}^{R,r}_C \tilde Y^{R,<}_C \tilde{g}^{R,a}_C  .
\end{split}
\end{equation}

The expression for $G^<_R$ can be obtained from Eq.~(\ref{eq:GlesserL}) by swapping
the index $L$ to $R$ and the self-energy $\tilde Y^{CR,x}_L$ by $\tilde Y^{CL,x}_R$
given in Eq.~(\ref{eq:Cregion_embed_pot1}).

\section{Current conservation condition}
\label{sec:currentconserv}

One of the most important physical properties that our formalism should
obey is the current conservation condition. Before deriving the current
conservation condition for the fully interacting system, we  briefly recall
the equivalent condition when the interactions are localised in the central
region only.

\subsection{Current conservation condition for the case of interaction only in the central region}
\label{app:currentconserv_nointerac}

Within the partitioned scheme devised by Meir and Wingreen,
i.e.\ interaction only in the central region $C$, it can be shown that
the following trace
\begin{equation}
\label{eq:app_CurrentConv1}
{\rm Tr}_n \left[ \Sigma^<(\omega) G^>(\omega) - \Sigma^>(\omega) G^<(\omega) \right] = 0
\end{equation}
vanishes for each $\omega$, where $\Sigma(\omega)$ is the total self-energy of the region $C$: 
$\Sigma=\Sigma^\MB_C + \Sigma^L_C + \Sigma^R_C$. The $\alpha$-lead's self-energy is
$\Sigma^\alpha_C= V_{C\alpha} g_\alpha V_{\alpha C}$.

Eq.(\ref{eq:app_CurrentConv1}) is derived from the definition 
$G^{>,<}(\omega) = G^r(\omega) \Sigma^{>,<}(\omega) G^a(\omega)$
in the region C, and hence 
$(G^r)^{-1}(G^>-G^<)(G^a)^{-1}=\Sigma^>-\Sigma^<=\Sigma^r-\Sigma^a=(G^a)^{-1}-(G^r)^{-1}$

With Eq.(\ref{eq:app_CurrentConv1}), we can derive a condition that must be fulfilled by
the interaction \SE \cite{Jauho:2005} in order to satisfy the
current conservation condition $I_L+I_R=0$ \cite{Jauho:2005}. This condition is given by:
\begin{equation}
\label{eq:app_CurrentConv2}
\int {\rm d}\omega\ {\rm Tr}_n \left[ \Sigma^{\MB <}_C\ G_C^> - \Sigma^{\MB >}_C\ G^<_C \right] = 0 ,
\end{equation}
which means that the integrated collision term must vanish. This is a condition familiarly
obtained from a Boltzmann-like treatment of scattering theory \cite{Jauho:2005}.
When the interaction \SE $\Sigma^\MB$ is derived from the so-called $\Phi$-derivable approximation
\cite{Baym:1962,vonBarth:2005,vanLeeuwen:2006,Kita:2010}, it automatically satisfies the condition 
given by Eq.(\ref{eq:app_CurrentConv2}). 

Eq.(\ref{eq:app_CurrentConv2}) can also be used as a measure of the accuracy of numerical schemes
used to calculate approximately the interaction \SE. It can also be used as a general constraint
equation in the determination a new functional forms for the interaction self-energy.

Now, when the interaction exists throughout the entire system, the derivation described above no 
longer holds and needs to be generalised to the presence of interaction within the leads and crossing 
at the left and right contacts.

\subsection{Current conservation condition for the case of interaction everywhere}
\label{app:currentconserv}

First we consider the general definition of the lesser and greater \GFs:
\begin{equation}
\label{eq:app_defGlessgrt_1}
G^< = 
( 1 + G^r \Sigma^r ) g^< ( 1 + \Sigma^a\ G^a ) + G^r\ \Sigma^<\ G^a .
\end{equation}
The first term represents the initial conditions $g^<$ before the interaction
and the coupling between the different regions are applied.

For the central region, we have chosen the initial condition such as 
$\langle n\vert g^< \vert m\rangle \equiv 0$ (see Eq.(\ref{eq:GlessC})). We could have
chosen another initial condition. Such choices have no effects on the steady-state 
regime when a steady current flow through the central region, however
the initial conditions play an important role in the transient behaviour of
the current \cite{Tran:2008,Myohanen:2008,Perfetto:2010,Velicky:2010}.

For the definition of the lesser left- and right-lead Green's functions, however,
it is  not possible
to neglect the initial conditions (before full interactions and coupling to the
region central are taken into account). It would not be physically correct to ignore
the presence of the left and right Fermi seas since they are the thermodynamical
limit of the two semi-infinite leads, and act as electron emitter and collector
in our model of a device.

By using the standard Dyson equations for $G^{a/r}$, we can recast 
Eq.(\ref{eq:app_defGlessgrt_1}) as
\begin{equation}
\label{eq:app_defGlessgrt_2}
G^< = G^r\ ( (g^r)^{-1} g^< (g^a)^{-1} + \Sigma^{<} )\ G^a
= G^r\ \bar{\Sigma}^{<}\ G^a ,
\end{equation}
with $\bar{\Sigma}^{<} = \Sigma^{<} + \gamma^<$ and $\gamma^< =
(g^r)^{-1} g^< (g^a)^{-1}$, and similarly for $G^>$.  Hence $\gamma^<
- \gamma^> = (g^a)^{-1} - (g^r)^{-1}$ and $\bar{\Sigma}^< -
\bar{\Sigma}^> = (G^a)^{-1} - (G^r)^{-1}$.
From these properties, it can be easily shown that 
\begin{equation}
\label{eq:app_collisionTerm_Tr_all}
{\rm Tr}_{\rm all} \left[ \bar\Sigma^< G^> - \bar\Sigma^> G^< \right] = 0 ,
\end{equation}
for each $\omega$. The trace runs over all indexes in the system ${\rm
  all} \equiv \{ \lambda, n, \rho \}$ and the interaction $\Sigma$ are
spread over the whole $L,C,R$ regions. This is a generalisation of
Eq.(\ref{eq:app_CurrentConv1}).

Because the trace runs over all the three subspaces, we can apply the
usual cyclic permutation and recast
Eq.(\ref{eq:app_collisionTerm_Tr_all}) as follows
\begin{equation}
\label{eq:app_collisionTerm_Tr_all_bis}
-{\rm Tr}_{\rm all} \left[ (\Sigma G)^< - (G \Sigma)^< \right] +
{\rm Tr}_{\rm all} \left[ \gamma^< G^> - \gamma^> G^< \right] = 0
\end{equation}
or equivalently
\begin{equation}
\label{eq:app_collisionTerm_Tr_all_ter}
\int {\rm d}\omega\ {\rm Tr}_{\rm all} \left[ (\Sigma G)^< - (G \Sigma)^< \right] +
{\rm Tr}_{\rm all} \left[ \gamma^> G^< - \gamma^< G^> \right] = 0
\end{equation}
Expanding the trace in the first term over each subspace ${\rm
  Tr}_{\rm all}[...] = {\rm Tr}_{\lambda}[..]+{\rm Tr}_{n}[...]+{\rm
  Tr}_{\rho}[...]$, one can easily identify the definition of the left $I_L$
and right $I_R$ currents from ${\rm Tr}_{\lambda}[...]$ and ${\rm
  Tr}_{\rho}[...]$ respectively (see Eq.(\ref{eq:ILafterEOM_omega})
above).

Hence the condition of current conservation $I_L+I_R=0$ leads to
\begin{equation}
\label{eq:curconserv_cond_1a}
\int {\rm d}\omega\ {\rm Tr}_n \left[ (\Sigma G)^< - (G \Sigma)^< \right] +
{\rm Tr}_{\rm all} \left[ \gamma^> G^< - \gamma^< G^> \right] = 0 .
\end{equation}

After further manipulation of the trace ${\rm Tr}_{n}[...]$ using the
rules of analytical continuation and the relationship between the
different \GFs and self-energies, we find that the current conservation implies
that
\begin{equation}
\label{eq:curconserv_cond_1b}
\int {\rm d}\omega\ {\rm Tr}_n 
\left[ (\Sigma^\MB_C + \tilde Y^{L+R}_C)^> G^<_C - (\Sigma^\MB_C +
  \tilde Y^{L+R}_C)^< G^>_C \right]  +
{\rm Tr}_{L,R} \left[ \gamma^> G^< - \gamma^< G^> \right] = 0 ,
\end{equation}
where in the second trace the sum runs only over the left and right
subspaces, because we have chosen the initial condition for the
central region C such that $\gamma^{>,<}=0$.

What is really interesting with the first trace ${\rm Tr}_{\rm n}[...]$
in Eq.(\ref{eq:curconserv_cond_1b}) is that it has exactly the same
form as Eq.(\ref{eq:app_collisionTerm_Tr_all}), but with all
quantities (as well as the trace) defined within the central region $C$
only 
(i.e.\ $\bar\Sigma \equiv \Sigma^\MB_C + \tilde Y^{L+R}_C$ in the region $C$).
By looking at the definition of $G^{>,<}_C$ given by Eq.(\ref{eq:GlessC}), 
we can establish that, equivalently to Eq.(\ref{eq:app_collisionTerm_Tr_all}),
the trace actually vanishes:
\begin{equation}
\label{eq:app_collisionTerm_Tr_C_only}
{\rm Tr}_n 
\left[ (\Sigma^\MB_C + \tilde Y^{L+R}_C)^> G^<_C - (\Sigma^\MB_C + \tilde Y^{L+R}_C)^< G^>_C \right]=0 ,
\end{equation}
for each $\omega$.
Therefore Eq.(\ref{eq:curconserv_cond_1b}) reduces to
\begin{equation}
\label{eq:curconserv_cond_1c}
\int {\rm d}\omega\ 
{\rm Tr}_{L,R} \left[ \gamma^> G^< - \gamma^< G^> \right] = 0 ,
\end{equation}
a condition which however is almost systematically verified (see \ref{app:gamma_proof}).

Hence it is better to consider Eq.(\ref{eq:app_collisionTerm_Tr_C_only})
to find the condition imposed by the current conservation.
Indeed by treating each contribution in Eq.(\ref{eq:app_collisionTerm_Tr_C_only})
separately and by integrating over the energy, 
we can introduce the definition of the currents $I_L$ and $I_R$ such as
\begin{equation}
\label{eq:curconserv_cond_2b}
\int {\rm d}\omega\ 
{\rm Tr}_n \left[ \Sigma^{\MB >}_C G^<_C - \Sigma^{\MB <}_C G^>_C \right]
-i_L(\omega) + \Delta i_L(\omega)
-i_R(\omega) + \Delta i_R(\omega)  = 0 ,
\end{equation}
where $I_\alpha= e/h \int i_\alpha(\omega) {\rm d}\omega$, and $\Delta
i_\alpha(\omega)= \left[ \tilde Y^{\alpha,>}_C G^<_C - \tilde Y^{\alpha,<}_C
  G^>_C \right] + i_L(\omega)$.

Hence the condition of current conservation $I_L+I_R=0$ leads to the final
important result of this section:
\begin{equation}
\label{eq:app_curconserv_cond_full}
\begin{split}
  \int {\rm d}\omega\
  & {\rm Tr}_{\alpha=L,C,R} \left[ \Sigma^{\MB >}_\alpha G^<_\alpha - \Sigma^{\MB <}_\alpha G^>_\alpha \right] \\
  & + {\rm Tr}_{n} \left[ (\tilde Y^{L,>}_C + \tilde\Upsilon^L_C -
    (\tilde\Upsilon^L_R)^\dag - \tilde\Upsilon^{L,l}_C) G^<_C  - (\tilde
    Y^{L,<}_C - \tilde\Upsilon^{L,l}_C) G^>_C +
    (\tilde\Upsilon^{L,l}_C + (\tilde\Upsilon^{L,l}_C)^\dag) G^a
  \right] \\
  & + {\rm Tr}_{n} \left[ \{L \leftrightarrow R \}
  \right] \\
  & = 0 .
\end{split}
\end{equation}

Using the properties of the $\tilde\Upsilon^L_C$ quantities (see  Section \ref{sec:theUpsilons}),
we can rewrite the condition imposed by the current conservation as
\begin{equation}
\label{eq:app_curconserv_cond_full2}
\begin{split}
  \int {\rm d}\omega\
  & {\rm Tr}_{\alpha=L,C,R} \left[ \Sigma^{\MB >}_\alpha G^<_\alpha - \Sigma^{\MB <}_\alpha G^>_\alpha \right] \\
  & + {\rm Tr}_{n} \left[ (\tilde Y^{L,>}_C - \tilde\Upsilon^{L,g}_C) G^<_C  - (\tilde
    Y^{L,<}_C - \tilde\Upsilon^{L,l}_C) G^>_C +
    (\tilde\Upsilon^{L,l}_C + (\tilde\Upsilon^{L,l}_C)^\dag) G^a
  \right] \\
  & + {\rm Tr}_{n} \left[ \{L \leftrightarrow R \}
  \right] \\
  & = 0 .
\end{split}
\end{equation}
To understand fully the conditions of current conservation, we make the following
observations:
\begin{itemize}
\item[(i)]
Eq.(\ref{eq:app_curconserv_cond_full2}) is the generalisation of
  Eq.(\ref{eq:app_CurrentConv2}) for the systems where the interaction
  spreads throughout. Like Eq.(\ref{eq:app_CurrentConv2}), it contains similar
  terms involving the left $L$ and right $R$ lead as well. But it also contains 
  terms arising from the fact that the
  interaction is crossing at the $LC$ and $RC$ contacts.  
  However, the physical interpretation of Eq.(\ref{eq:app_curconserv_cond_full2})
  still corresponds to the fact that the total integrated collision terms must
  vanish.
\item[(ii)]
For interaction present only within the $C$ region, one can show that
  $\tilde\Upsilon^{L,l}_C + (\tilde\Upsilon^{L,l}_C)^\dag=0$ as well as $\tilde
  Y^{L,<}_C - \tilde\Upsilon^{L,l}_C = 0$ and 
  $\tilde Y^{L,>}_C - \tilde\Upsilon^{L,g}_C = 0$ since $\Sigma^{\MB}_{LC}=V_{LC}$,
  and likewise for the terms involving the $RC$ interface.
  Furthermore, in that case, there are no interactions in the leads $\Sigma^{\MB}_{\alpha=L,R}=0$,
  and hence one recovers Eq.(\ref{eq:app_CurrentConv2}) as expected.
\item[(iii)] 
Eq.(\ref{eq:app_curconserv_cond_full2}) is also consistent with the
one of the main point made in Section \ref{sec:theUpsilons}: 
the current $I_\alpha$ ($\alpha=L$ or
  $R$) cannot be obtained from a trace over the central region defined
  as ${\rm Tr}_n [Y^{\alpha,<}_C G^>_C - Y^{\alpha,>}_C G^<_C]$.  
If it were the case, then the $\Delta i_\alpha(\omega)$ defined 
after Eq.~(\ref{eq:curconserv_cond_2b}) as
$\Delta
i_\alpha(\omega)= \left[ \tilde Y^{\alpha,>}_C G^<_C - \tilde Y^{\alpha,<}_C
  G^>_C \right] + i_L(\omega) \equiv 0$.
And the current conservation condition would
  imply that Eq.(\ref{eq:app_curconserv_cond_full2}) reduces to
\begin{equation}
\int {\rm d}\omega\ 
{\rm Tr}_{\alpha=\lambda,n,\rho} 
\left[ \Sigma^{\MB >}_\alpha G^<_\alpha - \Sigma^{\MB <}_\alpha G^>_\alpha \right] = 0.
\end{equation}
This equation concerns the collision terms within each of the three regions, but is not complete because
it does not show any constraint on the interaction crossing at the $LC$ and $RC$ contacts.

\end{itemize}

\section{Discussion and conclusion}
\label{sec:discuss}

In this paper, we have derived a complete and exact expression
  of the current crossing at the $LC$ (or $RC$) interface (defining the
  contact between the $L$ ($R$) lead and the central region $C$) for general
  systems with interaction both within each $L,C,R$ region and
  crossing at the left and right interfaces. Our result for the current
  Eq.(\ref{eq:ILfinal}) and Eq.(\ref{eq:IRfinal}) is general and obtained under only one
  approximation: there is no direct exchange and correlation effects between the
  left and right lead; a condition that is physically sound, especially
  for a large-ish central region where the spatial gap between the two
  electrodes is large enough so that the two electrodes interact 
  only indirectly via the central region.

Our formalism includes all the cases previously studied with any kind of interactions 
present in the central region only  
\cite{Mii:2003,Frederiksen:2004, Galperin:2004b,
  Mitra:2004, Pecchia:2004b, Chen_Z:2005, Ryndyk:2005, Sergueev:2005,
  Viljas:2005, Yamamoto:2005, Cresti:2006, Vega:2006, Zazunov:2007, Egger:2008, Dash:2010,
Thygesen:2007,Myohanen:2009,Dash:2011}.
It also include other classes of problems such as those where interactions also exist
within the leads but does not cross at the contacts \cite{Kletsov:2007}.
Our formalism also provides a natural way to extend cases where the excitations
exist in the leads and could cross at the contacts between the central region and the leads
\cite{Galperin:2006b,Y_Li:2011}.

We now discuss in more detail different open questions that are of importance for applications 
of our formalism to realistic systems.

\subsection{Location of the interfaces}

There is one arbitrary choice in our derivation: the location of
  the $LC$ and $RC$ interfaces with respect to the physical realistic
  system.  Such locations are somehow arbitrary in our formalism but
  could be conveniently chosen for practical numerical calculations.
  We have already seen that, in some cases, a local
  quasi-equilibrium is reached within the left and right leads. In such
  cases we
  get simplified results for the current, since the local
  non-equilibrium distribution functions $f^<_{L,R},f^{{\rm int}<}_{L,R}$ 
are equal to the local Fermi distribution in the left or
  right lead $f_{L,R}=f^{0 <}_{L,R}$.

\subsection{The extended molecule}

Our formalism provides a formal justification of the concept of
  the extended molecule that has been used so far within the
  conventional partitioned scheme with interaction present
  only in the $C$ region.  Being general, our formalism also provides the 
  corresponding  ``correction'' terms needed when the interactions cross at the
  contacts and when the contacts are not in their respective local
  (quasi) equilibrium.

Our work also justifies the scheme recently used in
  Ref.~\cite{Strange:2011} where two closed
  surfaces are used to define two sets of $LC$ and $RC$ interfaces around the
  molecule. Within the inner region, the electron-electron interaction is 
  calculated  via the $GW$ scheme with the many-body \SE
  $\Sigma(\tau,\tau')=G(\tau,\tau')W(\tau,\tau')$. While the
  screened Coulomb interaction $W=v+vPW$ and the polarization $P$ are calculated for the
  extended region in order to ensure a better treatment of non-local
  screening effects.  To some extent, this is an empirical way of
  defining the concept of generalised embedding potentials that we obtain
  in a formal manner in our formalism.

\subsection{Generalized embedding potential}

In a broader context, our formalism introduces in a formal
  manner the generalisation of the concept embedding potential to
  interacting cases.
  In the case originally studied by Meir and Wingreen, where the
  interaction is present only in the central region, the effects of
  the leads on the \GFs defined with the subspace of the region $C$ are
  taken into account via the so-called lead \SEs:
  $\Sigma^\alpha_C(\omega) = V_{C \alpha} g_\alpha(\omega) V_{\alpha
    C}$. These non-local \SEs are simply a matrix representation of
  the so-called embedding potential originally defined in real space
  by Inglesfield
  \cite{Inglesfield:1981,Fisher:1990,Ness:1997b,Inglesfield:2005,Inglesfield:2008}.
  The latter can be seen as defining a surface (or two interfaces)
  around the central region that the interaction does not
  cross. The non-locality of the embedding potentials arises only
  from the \GFs defined on this surface.

  In our formalism, when the interaction can cross the $LC/RC$
  interfaces, we obtain in a systematic way a generalisation of the
  embedding potentials $\tilde Y^\alpha_C$, defined 
  as $\tilde Y^\alpha_C(\omega) =
  \Sigma_{C\alpha}(\omega) \tilde g_{\alpha}(\omega) \Sigma_{\alpha
    C}(\omega)$.
  These generalised embedding potentials contain a ``double''
  non-locality, in the sense that the $\Sigma^\MB_{\alpha C}$ part of
  $\Sigma_{\alpha C}$ can have a larger spatial extent than the hopping
  matrix elements $V_{\alpha C}$.  Hence $\tilde Y^\alpha_C$ defines
  somehow not a simple surface but a ``buffer'' zone that is contained
  between two surfaces whose separation is related to the
  characteristic (spatial decay) length of the interaction \SEs
$\Sigma^\MB_{\alpha C} \equiv \Sigma^{\rm MB}(\vert \mathbf{x}_\alpha -\mathbf{x}_n \vert)$.

\subsection{Calculation of the \SEs}

For practical numerical calculations of the current, one needs to choose the form
for the interaction self-energies. They can be obtained from conventional many-body perturbation
theory and Feynman diagrammatics, extended onto the Keldysh contour.
The interaction \SE $\Sigma^\MB$ can be obtained from the $\Phi$-derivable 
conserving approximation
\cite{Baym:1962,vonBarth:2005,vanLeeuwen:2006,Kita:2010}, and then they should automatically satisfies 
the condition of current conservation.
We have given an example of such interaction \SEs in Ref.~\cite{Ness:2011} for
the case of electron-phonon coupling inside the region $C$ and crossing at the $LC$ interface.
One could also devise other functional forms for the self-energies, such as functional of the 
charge and spin densities, and or of the current density itself. 
In these cases, one should devise the
functionals such that the condition of current conservation is indeed fulfilled.

Finally, one should note that once the functional forms of the \SEs are chosen, one can 
perform the calculations self-consistently. The \SEs in the three regions $L,C,R$ and at
the interfaces $LC$ and $RC$ are functionals of the other electron (and phonon) \GFs
(or other physical quantities related to them such as the charge, spin or current density)
defined inside the $L,C,R$ regions as well as at the two $LC$ and $RC$ interfaces.
Hence the \GFs and \SEs need to be determined self-consistently in all the parts of the
system in order to get the current of a fully many-body interacting nanojunction at 
non-equilibrium.

\subsection{A special case for the \SEs}

When modelling the \SEs such that $\tilde\Upsilon^{L,l}_C+(\tilde\Upsilon^{L,l}_C)^\dag=0$,
we can express the current Eq.~(\ref{eq:ILfinal}) as
\begin{equation}
     \label{eq:IL_discuss}
\begin{split}
 I_L = \frac{e}{\hbar} \int \frac{d\omega}{2\pi} 
& {\rm Tr}_n\left[ (G^r_C-G^a_C) \tilde{\Upsilon}^{L,l}_C + G^<_C(\tilde\Upsilon^L_C -
  (\tilde{\Upsilon}^L_C)^\dagger) \right] \\
 + & {\rm Tr}_\lambda\left[\Sigma^{\MB,>}_{L} G^<_{L} 
     - \Sigma^{\MB,<}_{L} G^>_{L} \right] ,
\end{split}
\end{equation}
which bear even more resemblance to the Meir and Wingreen result, Eq.~(\ref{eq:IL_MeirWingreen}),
and hence could
be recast as a generalized Landauer-like expression for the current \cite{Ness:2010}.
In such cases, the condition of current conservation becomes
\begin{equation}
\label{eq:curconserv_cond_discuss}
\begin{split}
  \int {\rm d}\omega\
  & {\rm Tr}_{\alpha=L,C,R} \left[ \Sigma^{\MB >}_\alpha G^<_\alpha - \Sigma^{\MB <}_\alpha G^>_\alpha \right] \\
  & + {\rm Tr}_{n} \left[ (\tilde Y^{L,>}_C - \tilde\Upsilon^{L,g}_C) G^<_C  - (\tilde
    Y^{L,<}_C - \tilde\Upsilon^{L,l}_C) G^>_C  \right] \\
  & + {\rm Tr}_{n} \left[ \{L \leftrightarrow R \}
  \right] \\
  & = 0 .
\end{split}
\end{equation}
Eq.~(\ref{eq:curconserv_cond_discuss}) looks just like the sum of the integrated collision 
terms in the form of
\begin{equation}
  \int {\rm d}\omega\
  {\rm Tr}_{\beta} \left[ \Sigma^{{\rm int} >}_\beta G^<_\beta - \Sigma^{{\rm int} <}_\beta G^>_\beta 
  \right] = 0, 
\end{equation}
where $\Sigma^{\rm int}_\beta$ represent some interaction self-energy and the sum $\beta$ is over  
not only the three subspaces $L,C$ and $R$ but also the two interfaces $LC$ and $RC$.

By using the rules of analytical continuation and the relationships between the different \GFs,
one can find that, in general, the sum $\tilde\Upsilon^l_{LC}+(\tilde\Upsilon^l_{LC})^\dag$ is
given by 
\begin{equation}
\label{eq:sumUpsilon0}
\begin{split}
\tilde\Upsilon^{L,l}_C + (\tilde\Upsilon^{L,l}_C)^\dag  = 
 (\Sigma \tilde{g})^<_{CL}\  \Sigma^>_{LC}  - (\Sigma \tilde{g})^>_{CL}\  \Sigma^<_{LC}
+ \Sigma^>_{CL}\ (\tilde{g} \Sigma)^<_{LC} -  \Sigma^<_{CL}\ (\tilde{g} \Sigma)^>_{LC} .
\end{split}
\end{equation}
There are two different cases in which the sum $\tilde\Upsilon^{L,l}_C + (\tilde\Upsilon^{L,l}_C)^\dag$
vanishes:
\begin{itemize}
\item[(i)]
when the interaction is only present in central region. We have already discussed that case
in length in the paper.
\item[(ii)]
when interaction is instantaneous, i.e. local in time 
$\Sigma^{\rm MB}(\tau,\tau')=v^{\rm MB}(\tau)\delta(\tau-\tau')$. Hence there are
no lesser/greater components of the \SE $\Sigma^{{\rm MB},><}=0$. This occurs, as already
discussed, in mean-field based and in density-functional-based theories for
which a single (quasi-)particle description of the system is available.
\end{itemize}
Finally, it would be interesting to study and find cases which go beyond mean-field 
or density-functional-based methods, and
for which $\tilde\Upsilon^{L,l}_C + (\tilde\Upsilon^{L,l}_C)^\dag =0$. If such cases exist, 
it would still be possible to analyse their transport properties in terms of a
generalized Landauer-like approach \cite{Ness:2010}.

\ack
  We thank L. Kantorovich for useful comments, and K. Burke, P. Bokes,
R. Godby, M. Stankovski and M. Verstraete for stimulating discussions
on extended interaction.

\appendix

\section{Relationships between Green's functions}
\label{app:ccGFs}

By definition, complex conjugation of the different \GFs follows the rules:
\begin{equation}
\label{eq:app_cc}
\begin{split}
G^a(1,2)	& =   \left( G^r(2,1) \right)^*   \\
G^\gtrless(1,2)	& = - \left( G^\gtrless(2,1) \right)^* \nonumber
 \end{split}
\end{equation}
Similar expressions also hold for the \SEs $\Sigma$.

Furthermore, there exists relationships between the different components of the Green's
functions (or self-energies) on the Keldysh time-loop contour $C_K$. They are given
by:
\begin{equation}
\begin{split}
X^r = X^{++}-X^{+-} &= X^{-+}-X^{--} \\
X^a = X^{++}-X^{-+} &= X^{+-}-X^{--} \\
X^{++}+X^{--} &= X^{+-}+X^{-+}     \\ 
X^{-+}-X^{+-} &= X^r-X^a ,
\end{split}
\label{eq:app_gendef}
\end{equation}
with $X^{\eta_1 \eta_2}(12)\equiv G^{\eta_1 \eta_2}(12)$ or
$\Sigma^{\eta_1 \eta_2}(12)$, and where $(i=1,2)$ is the composite index for space-time
location $(\mathbf{x}_i,t_i)$ and $\eta_i$ is the index of the Keldysh time-loop contour 
$C_K$ branch ($+$ forward time arrow, $-$ backward time arrow) on which the time $t_i$ 
is located.
The conventional lesser and greater projections are defined respectively as
$X^< \equiv X^{+-}$ and $X^> \equiv X^{-+}$, and the usual
time-ordered (anti-time-ordered) as $X^t=X^{++}$ ($X^{\tilde t}=X^{--}$).

\section{Rules for analytical continuation}
\label{app:analyticalcontinuation}

For the following products $P_{(i)}(\tau,\tau')$ on the time loop contour $C_K$,
\begin{equation}
\label{eq:app_acontinuation_CK}
\begin{split}
P_{(2)}	& =  \int_{C_K} A B,   \\
P_{(3)}	& =  \int_{C_K} A B C,  \\
P_{(n)}	& =  \int_{C_K} A_1 A_2 ... A_n , \\ 		\nonumber
\end{split}
\end{equation}
we have the following rules for the different components  $P_{(i)}^x(t,t')$ on the real time axis:
$(x=r,a,>,<)$
\begin{equation}
\label{eq:app_acontinuation_t}
\begin{split}
P_{(2)}^\gtrless	& =  \int_t A^r B^\gtrless + A^\gtrless B^a,   \\ 
P_{(3)}^<	& =  \int_t A^< B^a C^a + A^r B^< C^a + A^r B^r C^<,  \\
P_{(n)}^r	& =  \int_t A_1^r A_2^r ... A_n^r  , \qquad
P_{(n)}^a	 =  \int_t A_1^a A_2^a ... A_n^a  .	\nonumber
\end{split}
\end{equation}

\section{Proof that $\int {\rm d}\omega\ {\rm Tr}_{L,R} \left[ \gamma^> G^< - \gamma^< G^> \right] = 0$}
\label{app:gamma_proof}

From the definition $\gamma^x(\omega)=(g^r)^{-1} g^x(\omega) (g^a)^{-1}$, the quantities
$\gamma^x$ have only matrix 
elements within the $\alpha=L,R$ lead since they involve only the non-interacting Green's functions.
Consequently we only have to prove that
\begin{equation}
\label{eq:app_curconserv_cond_2b}
\int {\rm d}\omega\ 
{\rm Tr}_{\alpha=L,R} \left[ \gamma^>_\alpha G^<_\alpha - \gamma^<_\alpha G^>_\alpha \right] = 0 .
\end{equation}

We now introduce the \NE distribution
$f^{0>,<}_\alpha(\omega)$ and $f^{>,<}_\alpha(\omega)$ defined, in the $\alpha$ lead subspace,
from the generalised Kadanoff-Baym ansatz \cite{Lipavski:1986} as follows:
\begin{equation}
\label{eq:app_def_NEdistrib}
\begin{split}
  G^x_\alpha(\omega) & = f^{0x}_\alpha(\omega) g^a_\alpha(\omega) - g^r_\alpha(\omega) f^{0x}_\alpha(\omega) ,\\
  G^x_\alpha(\omega) & = f^x_\alpha(\omega) G^a_\alpha(\omega) - G^r_\alpha(\omega) f^x_\alpha(\omega) ,
\end{split}
\end{equation}
with $x=>,<$. The distribution function $f^{0<}_\alpha$ is defined for the non-interacting and uncoupled
lead $\alpha$, and is given by the Fermi distribution of the lead $\alpha$ at its own equilibrium.  
Using these definitions, we can reformulate Eq.~(\ref{eq:app_curconserv_cond_2b}) after further
manipulation as 
\begin{equation}
\label{eq:app_curconserv_cond_trace}
\begin{split}
{\rm Tr}_{L,R} \left[ \gamma^> G^< - \gamma^< G^> \right] & = (2\pi)^2\
{\rm Tr}_{L,R} \left[ f^{0<}_\alpha B_\alpha A^G_\alpha -  f^<_\alpha A^G_\alpha B_\alpha  \right] \\
& = (2\pi)^2\ {\rm Tr}_{L,R} \left[ ( f^{0<}_\alpha - f^<_\alpha ) A^G_\alpha B_\alpha  \right] ,
\end{split}
\end{equation}
where
\begin{equation}
\label{eq:app_spectralfunc}
\begin{split}
A^G_\alpha(\omega) & = (G^r_\alpha(\omega) - G^a_\alpha(\omega))/2\pi {\rm i} , \\
B_\alpha(\omega) & = (g^a_\alpha)^{-1} A^g_\alpha (g^r_\alpha)^{-1} = 
((g^r_\alpha)^{-1} - (g^a_\alpha)^{-1})/2\pi {\rm i} .
\end{split}
\end{equation}
In the second equality of Eq.~(\ref{eq:app_curconserv_cond_trace}) we have used a diagonal
representation for the \NE distributions $f^{0<}_\alpha$ and $f^{<}_\alpha$.

There are several different cases for which Eq.~(\ref{eq:app_curconserv_cond_2b}) vanishes:
\begin{itemize}
\item[(i)]
Following the same reasoning as in Section \ref{sec:TrII}, when the $LC$ interfaces is located well
inside the lead $L$, the corresponding states are in their local equilibrium and 
$f^{0<}_\alpha - f^<_\alpha \sim 0$.
\item[(ii)]
Following the definition $B_\alpha=((g^r_\alpha)^{-1} - (g^a_\alpha)^{-1})/2\pi {\rm i}=
((\omega-H_L+{\rm i}\eta) - ( \omega-H_L-{\rm i}\eta))/2\pi {\rm i} = \eta/\pi \rightarrow 0$.
Therefore Eq.~(\ref{eq:app_curconserv_cond_trace}) vanishes unless $A^G_\alpha(\omega)$ is
singular at some energy. This would correspond to the appearance of localised bound states
in the $\alpha$ lead induced by the interaction (and/or by the \NE condition). Although
such an appearance cannot be ruled out in principle, it seems to correspond to pathological 
cases not relevant for the description of the metallic leads used in the experiments.
 \end{itemize}

\section*{References}

\end{document}